\documentclass[twocolumn,preprintnumbers,amsmath,amssymb]{revtex4}

\usepackage{graphicx}
\usepackage{dcolumn}
\usepackage{bm}
\usepackage{epsf}

\newcommand{\beq}{\begin{equation}}
\newcommand{\eeq}{\end{equation}}
\newcommand{\beqr}{\begin{eqnarray} \nonumber}
\newcommand{\eeqr}{\end{eqnarray}}
\newcommand{\beqrb}{\begin{eqnarray}}
\newcommand{\eeqrb}{\nonumber \end{eqnarray}}

\newcommand{\vect}[1]{\mathbf{#1}}

\newcommand{\fin}{\mbox{ .}}
\newcommand{\coma}{\mbox{ ,}}

\newcommand{\lrgspc}{\,\,\,\,\,\,\,\,\,}
\newcommand{\smlspc}{\,\,\,\,}
\newcommand{\rmvlrgspc}{\!\!\!\!\!\!\!\!\!}
\newcommand{\rmvsmlspc}{\!\!\!\!}

\begin{document}

\title{The spectrum of particles accelerated in relativistic, collisionless shocks}

\author{Uri Keshet and Eli Waxman}
\affiliation{%
Physics Faculty, Weizmann Institute, Rehovot 76100, Israel \\
keshet@wicc.weizmann.ac.il, waxman@wicc.weizmann.ac.il
}

\date{\today}

\begin{abstract}
We analytically study diffusive particle acceleration in relativistic, collisionless shocks. We find a simple relation between the spectral index $s$ and the anisotropy of the momentum distribution along the shock front. Based on this relation, we obtain $s=(3\beta_u-2\beta_u\beta_d^2+\beta_d^3)/(\beta_u-\beta_d)$ for isotropic diffusion, where $\beta_u$ ($\beta_d$) is the upstream (downstream) fluid velocity normalized to the speed of light. This result is in agreement with previous numerical determinations of $s$ for all $(\beta_u,\beta_d)$, and yields $s=38/9$ in the ultra-relativistic limit. The spectrum-anisotropy connection is useful for testing numerical studies and for constraining non-isotropic diffusion results. It implies that the spectrum is highly sensitive to the form of the diffusion function for particles travelling along the shock front. 
\end{abstract}

\maketitle

Diffusive (Fermi) acceleration of charged particles in collisionless shocks is believed to be the mechanism responsible for the production of non-thermal distributions of high energy particles in many astrophysical systems \cite{FermiAcc}. This process is believed to play an important role in, for example, planetary bow shocks within the solar wind, supernovae remnant shocks driven into the inter-stellar medium \cite{FermiAcc}, jets of radio galaxies \cite{Jets}, gamma-ray bursts (GRB's) \cite{grb}, and possibly shocks involved in the formation of the large scale structure of the universe \cite{Loeb00}. This phenomenon is common to shocks with widely differing physical characteristics (e.g. velocity and length scale), as evident from the above examples.

The Fermi acceleration process in shocks is still not understood from first principles (see, e.g., \cite{arons} for a discussion of alternative shock acceleration processes). Particle scattering in collisionless shocks is due to electro-magnetic waves. No present analysis self-consistently calculates the generation of these waves, the scattering of particles and their acceleration. Most analyses consider, instead, the evolution of the particle distribution adopting some Ansatz for the particle scattering mechanism (e.g. diffusion in pitch angle), and the "test particle" approximation, where modifications of shock properties due to the high energy particle distribution are neglected. 

This phenomenological approach proved successful in accounting for non-thermal particle distributions inferred from observations. The theory of diffusive particle acceleration in \emph{non-relativistic} shocks was first developed in 1977 \cite{Krymskii77, Axford77, Bell78, Blandford78}. Diffusive (Fermi) acceleration of test particles in non-relativistic shocks was shown, in particular, to lead to a power-law distribution of particle momenta, $d^3n/dp^3\propto p^{-s}$, with \cite{FermiAcc}
\beq s = 3\beta_u / (\beta_u-\beta_d) \fin \label{eq:SIsoNR} \eeq 
Here, $\beta_u$ ($\beta_d$) is the upstream (downstream) fluid velocity normalized to the speed of light. For strong shocks in an ideal gas of adiabatic index $\Gamma=5/3$, this implies $s=4$ (i.e. $p^2d^3n/dp^3\propto p^{-2}$), in agreement with observations.

Observations of GRB afterglows lead to the conclusion that the highly relativistic collisionless shocks involved produce a power-law distribution of high energy particles with $s=4.2\pm0.2$ \cite{grb_s}. This triggered a numerical investigation of particle acceleration in highly relativistic shocks \cite{Bednarz98}. The values of $s$ were calculated, in the "test particle" approximation and assuming velocity angle diffusion, for a wide range of Lorentz factors and several equations of state \cite[][and the references therein]{Bednarz98, Kirk00, Achterberg01}. In particular, $s$ was shown to approach the value $4.2$ for large Lorentz factors, in agreement with GRB observations.
The study of particle acceleration in relativistic shocks is of interest to many other high-energy astrophysical systems as well, e.g. jets in active galactic nuclei \cite{AGNjets} and in X-ray binaries (micro-quasars) \cite{microquasars}, and may be relevant for the production of ultra-high energy cosmic-rays \cite{uhecr}. 

The analysis of shock acceleration is more complicated in the relativistic case than in the non-relativistic case, mainly because the particle distribution near the shock is highly anisotropic. Due to this difficulty, only approximate numerical analyses (using Mote-Carlo simulations or eigenfunction methods) are available for relativistic shocks. In particular, an analytic expression for $s$ extending Eq.~(\ref{eq:SIsoNR}) to the relativistic regime is unavailable. In this paper we present an analytic study of diffusive particle acceleration in relativistic shocks, under the test particle and velocity angle diffusion approximations. 

\emph{I. Formalism.} Consider a shock front perpendicular to the z-axis, where the fluid flows in the positive z direction upstream ($z<0$) and downstream ($z>0$). For diffusion in the direction $\vect{\hat{p}}$ of rest frame momentum $p$, the stationary transport equation for particles with Lorenz factors well above the shock Lorenz factor, is \cite{Kirk87}
\beq \gamma_i(\beta_i+\mu_i) \partial_z f_i(\mu_i, p_i, z) = \partial_\mu j_i(\mu_i, p_i, z) \coma \label{eq:transport1} \eeq
where upstream/downstream indices $i\in\{u,d\}$ will be written only when necessary, $\gamma\equiv(1-\beta^2)^{-1/2}$ is the Lorenz factor, and $f(\mu,p,z)$ is the (Lorenz invariant) particle density in the phase space of $\mu\equiv \cos (\vect{\hat{p}}\cdot \vect{\hat{z}})$, $p$, and the distance from the shock front $z$ as measured in the \emph{shock frame}. The flux in momentum space is 
\beq j(\mu, p, z) = D_{\mu\mu}(\mu, p, z) \partial_\mu f(\mu, p, z) \coma \eeq
where $D_{\mu\mu}$ is the diffusion function, and $D_{\mu\mu}\propto (1-\mu^2)$ for isotropic diffusion. Assuming that $D_{\mu\mu}$ is separable in the form $D_1(\mu) D_2(p,z)$, Eq.~(\ref{eq:transport1}) may be written 
\beq (\beta+\mu) \partial_\tau f(\mu,p,\tau) = \partial_\mu [(1-\mu^2) D(\mu) \partial_\mu f(\mu,p,\tau)] \coma \label{eq:transport2} \eeq
where $D(\mu)\equiv D_1(\mu)/(1-\mu^2)$ and $\tau \equiv z \, D_2(p,z)/\gamma$. 

Next, we incorporate boundary conditions. Continuity across the shock front implies 
\beq f_u(\mu_u, p_u, \tau_u=0) = f_d(\mu_d, p_d, \tau_d=0) \coma \label{eq:continuity} \eeq 
where upstream and downstream quantities are related by a Lorenz boost of velocity $\beta_r=(\beta_u-\beta_d)/(1-\beta_u \beta_d)$;
$p_d=\gamma_r p_u(1+\beta_r \mu_u)$ and $\mu_d=(\mu_u+\beta_r)/(1+\beta_r \mu_u)$. 
Particle injection near the shock, and diffusion of particles into the far downstream only, imply that 
\beq \lim_{\tau \to -\infty} f_u = 0 \lrgspc \mbox{and} \lrgspc  \lim_{\tau \to +\infty} f_d = f_\infty \cdot p_d^{-s} \coma \label{eq:boundary} \eeq  
where $f_\infty$ is a constant.

The spectral index $s$ was previously calculated, by numerically finding some eigenfunctions of Eq.~(\ref{eq:transport2}) that satisfy the boundary conditions of Eq.~(\ref{eq:boundary}), and approximately matching the upstream and downstream solutions \cite{Kirk87,Heavens88,Kirk00}, or by Monte Carlo simulations \cite{Bednarz98, Achterberg01}. Ref. \cite{Achterberg01} has also exploited the relation \cite{Bell78}
\beq s= 3 - {{\ln\langle P_{ret}\rangle} \over {\ln\langle E_f/E_i \rangle}} \coma \label{eq:SvsPret} \eeq
where $\langle P_{ret}\rangle$ is the (flux averaged) probability of a particle crossing the shock downstream to return upstream, and $\langle E_f/E_i \rangle$ is the average energy gain per cycle. 
In the ultra-relativistic limit, $\beta_u=1$ and $\beta_d=1/3$, such studies have converged on the value $s=4.22\pm 0.02$ \cite{Bednarz98, Kirk00, Achterberg01}.

\emph{II. Anisotropy-Spectrum Connection.} The particle drift downstream implies that the distribution near the shock front is \emph{anisotropic}, more particles moving downstream than upstream. The distribution is nearly isotropic in the non-relativistic limit, where Eq.~(\ref{eq:SIsoNR}) is approximately recovered by demanding that $f_d(\tau_d=0)$ is isotropic \cite{Heavens88}. The lack of a characteristic momentum implies that the spectrum remains a power-law in the relativistic case, as verified numerically \cite{Bednarz98, Achterberg01}.

In a steady state, the particle distribution $f$ is stationary parallel to the shock front, implying that $\partial_\mu j=0$ for $\mu=-\beta$, as evident from Eq.~(\ref{eq:transport2}). Lead by this observation, we expand $f(\mu,p,\tau)$ and $D(\mu)$ around $\mu=-\beta$ (assuming smooth functions in this region, see e.g. \cite{Kirk00}):
\beqrb \label{eq:expansion} f_i(\mu_i,p_i,\tau_i) & = & \left[a_0^{(i)}(\tau_i) + a_1^{(i)}(\tau_i)(\mu_i+\beta_i) \right. \\ 
& & \left. \smlspc + a_2^{(i)}(\tau_i)(\mu_i+\beta_i)^2 + \cdots \right] p_i^{-s} \nonumber \\ 
D_i(\mu_i) = d_0^{(i)} & + & d_1^{(i)}(\mu_i+\beta_i) + d_2^{(i)}(\mu_i+\beta_i)^2+\cdots \eeqrb 
Using Eq.~(\ref{eq:continuity}), one can relate the coefficients $a_j(\tau=0)$ on both sides of the shock front. For the two lowest order terms, these relations yield an explicit expression for $s$;
\beq s = {{r_d-r_u} \over {\beta_u-\beta_d}} \coma \label{eq:SvsRud} \eeq
where we have defined 
\beq r \equiv \left. \gamma^{-2} {{a_1} \over {a_0}} \right|_{\tau=0} = 
\left. {j \over {D \, D_2 \, f}} \right|_{\tau=0,\mu=-\beta} \fin \eeq 

The physical significance of $a_1$ and $a_0$ may be demonstrated as follows. For rest frame velocity angle diffusion, the mass and momentum flux are generally \emph{not} conserved, and particle energy is conserved only in the rest frame. Conservation of particle number and rest frame energy yield the one-dimensional continuity equation, 
\beq \partial_{t^\prime} \int_{-1}^1 \, d\mu \,f(\mu,p,z) + \partial_{z^\prime} \int_{-1}^1 \, d\mu \,\mu \,f(\mu,p,z) = 0 \coma \eeq 
where $t^\prime$ and $z^\prime$ are rest frame variables. Transforming to shock frame variables, we find that in a steady state 
\beq \int_{-1}^1  \, d\mu (\beta+\mu) f(\mu,p,z) = C p^{-s} \coma \label{eq:Gconserved} \eeq 
where $C$ is a constant determined by the boundary conditions; $C_u=0$ upstream and $C_d=2\beta_d f_\infty>0$ downstream \cite{Note1}. If we define the convection towards the upstream and towards the downstream as 
\beq G_+ \equiv \int_{-\beta}^{1} \,d\mu \,g \lrgspc \mbox{and} \lrgspc 
G_- \equiv \int_{-1}^{-\beta} \,d\mu \,(-g) \coma \eeq
then $G_+$ and $G_-$ are non-negative, $G_+=G_-$ upstream, and $G_+=G_-+C_d$ downstream. In addition, 
\beq \partial_\tau G_+ = \partial_\tau G_- = - a_1(\tau) D(\mu=-\beta) \gamma^{-2} \coma \label{eq:GpGm} \eeq 
reflecting the fact that as particles turn around from heading upstream to heading downstream or vice versa, they must diffuse through a state where they propagate parallel to the shock front. 
Finally, $a_0 = f (\mu=-\beta)$ is the normalization of $f$ along the shock front. 

Next, we exploit the stationary particle distribution parallel to the shock front. Substituting Eq.~(\ref{eq:expansion}) into Eq.~(\ref{eq:transport2}) for $\mu=-\beta$, we find a simple connection between the coefficients $a_1$ and $a_2$, valid for any $\tau$:
\beq a_2(\tau)=-a_1(\tau)\left(\beta \gamma^2+{1 \over 2}{d_1 \over d_0}\right) \fin \label{eq:coeff} \eeq 
When extrapolated to $\tau=0$ on both sides of the shock front, this result may be combined with the relation between $a_2^{(u)}$ and $a_2^{(d)}$ (cf. Eq.~[\ref{eq:continuity}] and Eq.~[\ref{eq:expansion}]) and with Eq.~(\ref{eq:SvsRud}), to yield a relation between $r_u$ and $r_d$
\beq r_u(r_u + d_u + \beta_u + \beta_d) = r_d(r_d + d_d + \beta_u + \beta_d) \coma \label{eq:RUvsRD} \eeq 
where $d_i \equiv \gamma_i^{-2} d_1^{(i)}/d_0^{(i)}$ is a measure of the deviation from isotropic diffusion, for particles moving almost parallel to the shock front. Alternatively, Eq.~(\ref{eq:SvsRud}) and Eq.~(\ref{eq:RUvsRD}) may be combined to yield expressions for $s$ as a function of $r_i$ on only one side of the shock, 
\beqrb s(r_d) & = & {1 \over {2(\beta_u-\beta_d)}} \left[ \right. \beta_u+\beta_d+2r_d+d_u \pm \label{eq:SvsRd} \\ 
& & \left. \sqrt{(\beta_u+\beta_d+2r_d+d_u)^2 + 4r_d(d_d-d_u)} \right] \eeqrb
and 
\beqrb s(r_u) & = & {1 \over {2(\beta_u-\beta_d)}} \left[ \right. -(\beta_u+\beta_d+2r_u+d_d) \pm \label{eq:SvsRu} \\ 
& & \left. \sqrt{(\beta_u+\beta_d+2r_u+d_d)^2 + 4r_u(d_u-d_d)} \right]
\fin \eeqrb 
For isotropic diffusion, where $D_i(\mu)$ is constant, $d_i=0$, Eq.~(\ref{eq:RUvsRD}) simplifies to $r_u+r_d+\beta_u+\beta_d=0$, and 
\beq s = {{2 r_d + \beta_u + \beta_d} \over {\beta_u - \beta_d}} = {{(-2 r_u) - \beta_u - \beta_d} \over {\beta_u - \beta_d}} \fin \label{eq:SvsRduIso} \eeq

The above analysis can be repeated by expanding the distribution function in any other frame. For example, expansion around $\mu_s\approx 0$ in the shock frame yields 
\beqrb s(r_s) & = & {1 \over 2} + {r_s \over {\beta_u+\beta_d}}+\widetilde{d}_u\beta_u-\widetilde{d}_d\beta_d \pm \label{eq:SvsRs} \\ & & \rmvlrgspc \rmvsmlspc 
\sqrt{ \left( {1\over2} + {r_s \over {\beta_u+\beta_d} } +\widetilde{d}_u\beta_u-\widetilde{d}_d\beta_d \right)^2 + 2r_s(\widetilde{d}_d-\widetilde{d}_u)} \coma \eeqrb 
where $\widetilde{d}_i\equiv d_i/(2\beta_u^2-2\beta_d^2)$, and $r_s\equiv a_1^{(s)}/a_0^{(s)}$ is related to $r_u$ and $r_d$ through 
$r_s = (\beta_u r_d - \beta_d r_u)/(\beta_u - \beta_d)$.
For isotropic diffusion, Eq.~(\ref{eq:SvsRs}) simplifies to 
\beq s = 1 + 2 r_s / (\beta_u + \beta_d) \label{eq:SvsRsIso} \fin \eeq 

\emph{III. Analytic expression for $s$.} 
The boundary conditions at $\tau \rightarrow \pm \infty$ were not used in the analysis leading to Eqs.~(\ref{eq:SvsRd}-\ref{eq:SvsRsIso}), but are needed in order to determine the anisotropy parameters $a_1/a_0$. In general, the convection gradually decreases at larger distances from the shock front. Eq.~(\ref{eq:GpGm}) thus implies that $a_1^{(u)}>0$ and $a_1^{(d)}<0$. For isotropic diffusion, this yields the constraint 
\beq s_{iso} > (\beta_u+\beta_d) / (\beta_u-\beta_d) \fin \eeq 

More important consequences of the above connection derive from cases where $s$ is known; (i) the non-relativistic limit of Eq.~(\ref{eq:SIsoNR}); and (ii) the limit of infinite compressibility $\beta_d=0$, where $\langle P_{ret} \rangle = 1$ and thus  
\beq s(\beta_u>0,\beta_d=0)=3 \fin \label{eq:Scompress} \eeq 
For isotropic diffusion, Eqs.~(\ref{eq:SIsoNR}), (\ref{eq:SvsRduIso}) and (\ref{eq:Scompress}) imply 
\beqrb a_1^{(d)} / a_0^{(d)} & = & \beta_u - {1\over 2} \beta_d + \beta_d A_d(\beta_u,\beta_d) \label{eq:RuRdIso} \lrgspc \mbox{and} \\ 
a_1^{(u)} / a_0^{(u)} & = & -\gamma_u^2 \left[ 2\beta_u + {1\over 2} \beta_d + \beta_d A_u(\beta_u,\beta_d) \right] \fin \eeqrb 
The unknown functions $A_i$ are at least second order, even \cite{Note2} functions of $\beta_u$ and $\beta_d$. Hence, in the non-relativistic limit, 
$a_1^{(d)}/a_0^{(d)} \simeq \beta_u - \beta_d/2$ and $a_1^{(u)}/a_0^{(u)} \simeq -2 \beta_u- \beta_d/2$. 

\begin{figure}[h]
\centerline{\epsfxsize=7.9cm \epsfbox{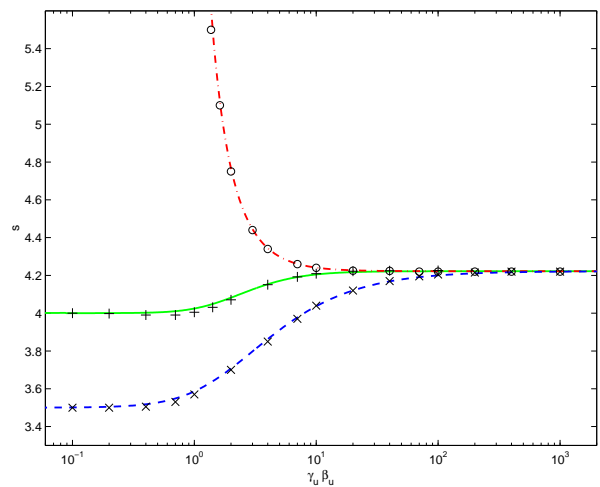}} 
\caption{\label{fig:SIso} Comparison between Eq.~(\ref{eq:SIso}) (curves) and previous numerical results \cite{Kirk00} (symbols). The spectral index $s$ is shown as a function of $\beta_u \gamma_u$ for three different types of shocks; a strong shock with the J\"{u}ttner/Synge equation of state (solid curve vs. crosses), a strong shock with fixed adiabatic index $\Gamma=4/3$ (dashed curve vs. x-marks), and a shock in a relativistic gas where $\beta_u \beta_d=1/3$ (dash-dotted curve vs. circles). For a description of these equations of state, see Ref. \cite{KirkDuffy99}. 
}
\end{figure}

In the relativistic case, the shock front distribution is far more isotropic in the downstream frame than it is in the upstream or the shock frames. In particular, $a_1/a_0$ includes (at least) third order terms (in $\beta_u$ or $\beta_d$) upstream, but could be of first order downstream. Using only the first order terms of $a_1/a_0$ downstream (i.e. assuming $A_d=0$) yields a simple expression for $s$, 
\beq s_{iso} = (3\beta_u - 2\beta_u \beta_d^2 + \beta_d^3) / (\beta_u - \beta_d) \coma \label{eq:SIso} \eeq
which is in excellent agreement with numerical studies \cite{Kirk00, Achterberg01} over the entire relevant range of $\beta_u$ and $\beta_d$. This is demonstrated in Figure \ref{fig:SIso}, for three different equations of state. In the ultra-relativistic limit, Eq.~(\ref{eq:SIso}) implies 
\beq s_{iso}(\beta_u\rightarrow 1, \beta_d \rightarrow 1/3) = 38/9 = 4.222\ldots \coma \eeq 
in excellent agreement with Refs. \cite{Bednarz98, Kirk00, Achterberg01}. The expression $a_1^{(d)}/a_0^{(d)}=\beta_u-\beta_d/2$ itself agrees with the downstream distribution calculated numerically \cite[][Figure 3]{Kirk00} even in the ultra-relativistic case, although we can not prove that this first order expansion is sufficient. 

\emph{IV. Other consequences.}
Eqs.~(\ref{eq:SvsRd}-\ref{eq:SvsRsIso}) directly relate the spectral index to the distribution of particles travelling nearly along the shock front. Note that Eqs. (\ref{eq:SvsRd}), (\ref{eq:SvsRu}) and (\ref{eq:SvsRs}) hold for any form of the diffusion function. These results provide an important consistency check for numerical methods such as used in Refs. \cite{Kirk87, Heavens88, Bednarz98, Kirk00, Achterberg01}. 

In some cases, $a_1/a_0$ can be read directly from the calculated particle distribution. For example, Figure 3 of Ref. \cite{Kirk00} shows the shock front distribution in both downstream and shock frames, for isotropic diffusion in an ultra-relativistic shock with $(\beta_u=0.995,\beta_d=0.328$). The figure yields $a_1^{(d)}/a_0^{(d)}\simeq 0.85$ and $a_1^{(s)}/a_0^{(s)}\simeq 2.1$, corresponding to $s\simeq 4.2$ according to both Eq. (\ref{eq:SvsRduIso}) and Eq.~(\ref{eq:SvsRsIso}). For the same shock and non-isotropic diffusion with $D(\mu)\propto (\mu^2+0.01)^{-1/2}$, Figure 5b of Ref. \cite{Kirk00} implies that $a_1^{(s)}/a_0^{(s)}\simeq 1.9$. Using Eq. (\ref{eq:SvsRs}), this also corresponds to $s \simeq 4.2$. 

The spectrum-anisotropy connection may be used to associate each eigenfunction $k$ of Eq.~(\ref{eq:transport2}) with its corresponding spectral index $s_k$. For isotropic diffusion in the ultra-relativistic limit, the shock front distribution coincides up to a $10\%$ accuracy with the first upstream eigenfunction $f_1$ \cite{Kirk00}, where in the shock frame $f_1 \propto (1-\beta_u \mu_s)^{-s}\exp[-(1+\mu_s)/(1-\beta_u\mu_s)]$. According to Eq.~(\ref{eq:SvsRsIso}), this gives $s_1 = (2+\beta_u-\beta_d) / (\beta_u-\beta_d) = 4$ in the ultra-relativistic limit. 

For non-isotropic diffusion, Eqs.~(\ref{eq:SvsRd}) and (\ref{eq:Scompress}) imply that, to lowest order in $\beta_d$, 
\beq a_1^{(d)} / a_0^{(d)} \simeq \beta_u {{\beta_u-\beta_d/2-d_u/2} \over {\beta_u+(d_d-d_u)/6}} \fin \label{eq:RdNonIso} \eeq 
For the non-isotropic diffusion function solved numerically \cite{Kirk00}, $D(\mu)\propto (\mu^2+0.01)^{-1/2}$, the approximation Eq.~(\ref{eq:RdNonIso}) is in poor agreement with the numerical data, although in the ultra-relativistic limit it yields $s=4.26$, similar to the result $s=4.21$ of Ref. \cite{Kirk00}. 

It is interesting to note that Eq.~(\ref{eq:SIso}) may be written in the form $s_{iso} = 3 - (-3\beta_d + 2\beta_u \beta_d^2 - \beta_d^3)/(\beta_u - \beta_d)$. Comparing this result with Eq.~(\ref{eq:SvsPret}) may suggest that 
\beqr \langle P_{ret} \rangle = \exp \left(-3\beta_d + 2\beta_u \beta_d^2 - \beta_d^3\right) \\
\mbox{and} \lrgspc \langle E_f/E_i \rangle = \exp \left(\beta_u - \beta_d \right) \fin \label{eq:guess} \eeqr 
Although other choices that conserve $\ln\langle P_{ret} \rangle/\ln\langle E_f/E_i \rangle$ are possible, Eq.~(\ref{eq:guess}) does agree well with numerical studies of the ultra-relativistic limit, where $\langle P_{ret} \rangle=0.439\pm0.007$ and $\langle E_f/E_i \rangle=1.97\pm0.02$ \cite{Achterberg01}. 

\emph{V. Summary.} We have analytically studied the spectrum of test particles accelerated by an arbitrary relativistic shock in the diffusion limit. A simple relation was shown [Eqs.~(\ref{eq:SvsRd})-(\ref{eq:SvsRsIso})] to exist between the spectral index $s$ and a measure of the particle anisotropy, $a_1/a_0$. For isotropic diffusion, the lowest order terms in $a_1/a_0$ downstream yield an expression for $s$ [Eq.~(\ref{eq:SvsRduIso})], which is in excellent agreement with previous numerical studies (Figure \ref{fig:SIso}) over the entire relevant range of $\beta_u$ and $\beta_d$. In the ultra-relativistic limit, it yields $s=38/9$, in agreement with previous studies and with GRB observations.  

The spectrum-anisotropy connection in Eqs. (\ref{eq:SvsRd}), (\ref{eq:SvsRu}) and (\ref{eq:SvsRs}) holds for any diffusion function $D(\mu)$. It indicates that $s$ is particularly sensitive to the form (first derivative) of $D(\mu)$ for upstream and downstream particles travelling along the shock front. 
This connection is also independent of the test particle approximation, providing a useful tool or at least a consistency check for various studies in the diffusion limit, including a future self-consistent calculation of particle scattering, acceleration, and electro-magnetic wave generation. 

This work was supported by Minerva and ISF grants. 
We are grateful to Amir Sagiv for helpful discussions.

\end{document}